\documentclass[11pt,noheadings]{amsart}

\usepackage{geometry}                
\usepackage{graphicx}
\usepackage{amssymb}
\usepackage{epstopdf}
\title{About the importance of the nuclear recoil in $	\alpha$ emission near the DNA}

\author{E. Lodi Rizzini, A. Bianconi, M. Corradini, M.Leali, V. Mascagna, \\L. Venturelli and N. Zurlo\\
Dipartimento di Chimica e Fisica per l'Ingegneria e per i Materiali,
 Universit\`a di Brescia, 25123 Brescia, Italy \\
Istituto Nazionale di Fisica Nucleare, Gruppo Collegato di Brescia, 25123 Brescia, Italy} 
\begin{document}

\begin{abstract}
The effect of the energy deposition inside the human body made by radioactive substances
is discussed. For the first time, we stress the importance of the recoiling nucleus in such reactions,
particularly concerning the damage caused on the DNA structure.
\end{abstract}

\maketitle

Since the discovery of the radioactivity, one of the most important scientific issues has been the
evaluation  of the consequences for the human health of the contact or the ingestion of substances
having such a property.

With this aim, it has been first investigated how a nucleus can release energy to the exterior
through processes which are non chemical but still due to electromagnetic forces.
The ``nuclear'' origin of the physical processes involved is so that
the energy released in a single event is several orders of magnitude larger 
than the ones released during usual chemical processes concerning our life.
The different processes involving the emission of a high-energy photon ($\gamma$ ray),
of an electron or a positron ($\beta^+$ or $\beta^-$ ray) or a helium  nucleus (made by 2 protons and two 
neutrons, $\alpha$ ray) have been identified. Another recurrent possibility is the spontaneous neutron 
emission in fissile nuclei. 

The biological hazardousness of these processes has been classified, with special care to
their effect on the DNA damage. Consequently the possible damage has been evaluated
according to the energy released during the interaction with the organic matter, 
and namely with the DNA.

Anyway in the current literature there is no focus on the object having the larger density of 
energy released in the matter: the nucleus recoiling after the $\alpha$ particle emission.
The momentum conservation imposes  that the $\alpha$ particle momentum equals the momentum 
carried away by the resulting nucleus. Then, the kinetic energy of the nucleus and the $\alpha$ particle
are univocally linked for each process by the ratio between their masses, which is of the order of 60.  

As a consequence in the case of Po$^{212}$, emitting an $\alpha$ particle with 8748 keV,
a Pb$^{208}$ nucleus is produced recoiling with about 170 keV kinetic energy. In the case of
Th$^{232}$,  emitting an $\alpha$ particle with 4012 keV, the nucleus of Ra$^{228}$ has a 
kinetic energy of 66 keV. These are two limiting cases for the $\alpha$ particle emitted by many 
high-Z  nuclides.

If we assume that the recoiling nucleus is totally ionized, these kinetic energies 
imply a high energy transfer to the molecules and to the atoms encountered along the path 
before the nucleus stops, as it is well known. The electrons 
(either  lost by the recoiling nucleus or produced along the path by ionization) are essential to evaluate the damage caused to the tissues around the decaying 
nucleus \cite{science2000}.

If the free path of the $\alpha$ particle can be estimated in some tens of micrometers in the cell 
tissue, i.e. of the same order of magnitude of the cell size,  the path of the recoiling nucleus will be
much shorter, of the order of hundreds of nanometers or less. The nucleus recoiling after an $\alpha$ 
particle emission will give rise to an energy deposition  (on a DNA structure nearby) even two orders of 
magnitude larger than the  $\alpha$ particle itself. The evaluation of the relative probability will be done 
in forthcoming and more specific articles.

Important differences emerge when we take into account the different decay chains,  recalling that
the daughter nucleus  will travel only for some hundreds of nanometers, and consequently it will stay 
near the DNA, if the parent nucleus was. It is well known the thorotrast case and the long-term 
induced mortality (even after decades) \cite{thorotrast-germany} .
 It is useful to point out that six $\alpha$ decays occurs, 
starting from Th$^{232}$ and bringing to Pb$^{208}$, only the first two of them 
Th$^{232} \rightarrow$ Ra$^{228}$, (halftime = 5,75 years)  and Th$^{228} \rightarrow$ Ra$^{224}$  (halftime = 1.9 years), have timescales of the order of one year . The timescales of the subsequent
decays are much shorter, spanning from tens of hours to a fraction of a second.

The  U$^{238}$ chain is interrupted in the human tissue after the first $\alpha$ decay  
U$^{238} \rightarrow$ Th$^{234}$ by the subsequent $\beta^-$ emission producing
Pa$^{234}$ and U$^{234}$, having a half-life of 246,000 years. Also the U$^{235}$ chain is similarly
interrupted.  It is straightforward to understand in this way the different hazardousness of
the different chains for the DNA structure.

It is not discussed here the hazardousness implied by the size of the possible aggregates 
of atoms or molecules containing radionuclides $\alpha$ emitters. There is anyway a vaste litterature
about the solubility of these substances and their distribution in the human body \cite{thorotrast-liver,thorium-inhaled}.
Also the contribution of the low-energy electrons emitted by the recoiling nucleus during his motion 
must be carefully considered \cite{science2000}, but it cannot be discussed in the present article
(which obviously is not including all the observations and the consequences of this effect).

In the end, we mention that also the neutron emission can produce the recoil of the nucleus,
with kinetic energies reduced by a factor of $\approx$ 4 with respect to the $\alpha$ case
(because of the mass ratio). However, the neutron energy is typically much lower than
the $\alpha$  particle energy.

\end{document}